\documentclass[]{aastex6}

\begin{document}

\title{Preconditioning of interplanetary space due to transient CME disturbances}
\author{M.~Temmer\altaffilmark{1}, M.A.~Reiss\altaffilmark{1}, L.~Nikolic\altaffilmark{2}, S.J.~Hofmeister\altaffilmark{1}, and A.M.~Veronig\altaffilmark{1}}
\email{manuela.temmer@uni-graz.at}

\altaffiltext{1}{Institute of Physics, University of Graz, Universit\"atsplatz 5/II, 8010 Graz, Austria.}
\altaffiltext{2}{Canadian Hazards Information Service, Natural Resources Canada, Ottawa, Canada.}

\begin{abstract}
Interplanetary space is characteristically structured mainly by high-speed solar wind streams emanating from coronal holes and transient disturbances such as coronal mass ejections (CMEs). While high-speed solar wind streams pose a continuous outflow, CMEs abruptly disrupt the rather steady structure causing large deviations from the quiet solar wind conditions. For the first time, we give a quantification of the duration of disturbed conditions (preconditioning) for interplanetary space caused by CMEs. To this aim, we investigate the plasma speed component of the solar wind and the impact of \textit{in situ} detected CMEs (ICMEs), compared to different background solar wind models (ESWF, WSA, persistence model) for the time range 2011--2015. We quantify in terms of standard error measures the deviations between modeled background solar wind speed and observed solar wind speed. Using the mean absolute error, we obtain an average deviation for quiet solar activity within a range of 75.1–-83.1 km/s. Compared to this baseline level, periods within the ICME interval showed an increase of 18--32\% above the expected background and the period of 2 days after the ICME displayed an increase of 9--24\%. We obtain a total duration of enhanced deviations over about 3 and up to 6~days after the ICME start, which is much longer than the average duration of an ICME disturbance itself ($\sim$1.3~days), concluding that interplanetary space needs $\sim$2--5~days to recover from the impact of ICMEs. The obtained results have strong implications for studying CME propagation behavior and also for space weather forecasting. 
\end{abstract}

\keywords{Sun: coronal mass ejections (CMEs) --- solar wind --- Sun: heliosphere --- solar-terrestrial relations}

\section{Introduction}

The solar wind during quiet Sun conditions consists of a continuous outflow of slow solar wind streams from the low-latitude streamer belt and fast solar wind stemming from coronal holes. This pattern of alternate slow and fast solar wind streams repeats itself with each solar rotation. It is disrupted by transient events, such as coronal mass ejections (CMEs), which present dynamic regions of reduced plasma density, enhanced speed and radially stretched magnetic field, therefore causing strong changes in the solar wind properties on much shorter timescales \citep[see e.g.,][]{schwenn06}. Such disturbed conditions of interplanetary space may be labeled as ``preconditioning'' for subsequent CMEs expelled into these regions of interplanetary space. As CMEs evolve from the Sun, they are initially governed by the Lorentz force and in their later evolution phase by the aerodynamic drag force \citep[see e.g.,][]{cargill96}. The drag force becomes dominant as the CME propagates into interplanetary space and is usually reflected in the deceleration of CMEs which are faster than the ambient solar wind speed and acceleration of CMEs which are slower than the solar wind \citep{gopalswamy00,vrsnak01}. 

Keeping in mind that the average occurrence frequency of CMEs is about 0.5 per day during the solar minimum phase and 2--3 per day during solar maximum \citep{stcyr00}, where successive CMEs are often launched from the same active region, we may assume that a continuous, i.e.,\,non-disturbed, solar wind flow towards a certain direction prevails only during times of low solar activity. This is also reflected in the performance of background solar wind models. \cite{reiss16} showed for different (semi-)empirical background solar wind models, that the correlation coefficient between modeled and observed solar wind speed at Earth is higher for times without CMEs \citep[see also][]{kohutova16}. Even sophisticated numerical MHD background solar wind models tend to fail during times of enhanced solar activity, while they are well matching the actual observation during times of low activity \citep[e.g.,][]{jian11,gressl14}. 

The conditions of interplanetary space, i.e., the ambient magnetic field and plasma flow, in which CMEs are embedded in, affect the CME transit times and their geo-effectiveness. Preconditioning of interplanetary space due to transient disturbances has immediate consequences for the propagation behavior of CMEs as was, e.g., impressively revealed by the event from July 23, 2012. This CME was observed to be super-fast and traversed the distance Sun--STEREO-A in less than 21~hours due to the changed conditions of interplanetary space most probably caused by the CME launched 3.5 days earlier from the same active region \citep{liu14,temmer15}. If Earth-directed, this event would have been one of the most extreme Space Weather events with an estimated $Dst>$1100~nT \citep{baker13}. Thus, the question arises how long does the interplanetary space need to recover from CME disruptions?

In this study we will for the first time quantify the duration of preconditioned phases of interplanetary space as a consequence of preceding CMEs. To this aim, we apply background solar wind models and compare the modeled solar wind speed to actual observations for times with and without CMEs. We investigate the time range 2011--2015, which is favorable as solar cycle 24 is rather weak and we may expect extended periods of undisturbed solar wind flow. \textit{In situ} measurements show for time ranges influenced by interplanetary CME (ICME) disturbances on average plasma speeds clearly above the background level in addition to significant signatures of compressed sheath plasma behind shocks and a smooth magnetic field inside the ejecta clouds \citep[e.g.,][]{schwenn06}. We will show that these preconditioned periods prevail for much longer time ranges than the duration of the ICME disturbance itself. In fact, we find that the impact of CMEs affects interplanetary space over periods of about 3 and up to 6~days. This has important implications for CME propagation models and space weather forecasting algorithms, as precursor events over a period of up to 6~days may need to be taken into account. 

\section{Data and methods}\label{methods}

We investigate the effect of in-situ detected CMEs (ICMEs) on the interplanetary solar wind speed profile for the time span 2011--2015. We use level-2 solar wind data (solar wind bulk speed [km/s]) from the ACE/SWEPAM instrument located at L1 \citep{mccomas98}. The 64-second resolution solar wind data were linearly interpolated to a 4-min time resolution. The arrival times and mean impact speeds of the ICMEs (shock-sheath, magnetic cloud structure) are extracted from the Richardson and Cane list \citep[``R\&C List''\footnote{\url{http://www.srl.caltech.edu/ACE/ASC/DATA/level3/icmetable2.htm}}][]{richardson10}.

To simulate the continuous, quiet Sun background solar wind speed (without CME disturbances), we use the empirical solar wind forecast (ESWF) and the Wang-Sheeley-Arge (WSA) model. The ESWF and WSA model results are given for Earth distance, and cover a cadence of 6~h (hour) and 4~h, respectively. For comparison, all data used for the present study are linearly interpolated onto a 6~h time grid. The ESWF models the quiet solar wind speed at 1~AU based on the fractional area of coronal holes detected in EUV filtergrams at the central meridian of the Sun covering a slice of 15$^{\circ}$ longitudinal width corresponding to the solar rotation within approximately 1 day \citep[see][]{vrsnak07,rotter12}\footnote{An operational version of ESWF, used for real-time forecasting of the solar wind speed at Earth, is presented at \url{swe.uni-graz.at} under ``Services".}. The coronal hole areas are extracted using an intensity-based threshold technique applied on SDO/AIA image data in the 19.3~nm band with a time cadence of 6~hours \citep[see][]{reiss16}. In contrast, the WSA model empirically relates the solar wind speed to the flux-tube expansion factor of open magnetic field lines \citep{arge00,arge03} which are computed from the potential-field source-surface model \citep[PFSS;][]{altschuler69} applied on magnetograms from the Global Oscillation Network Group (GONG). For more details on the application of this model to derive the background solar wind speed at 1~AU, we refer to \cite{nikolic14}. As a cross-check, we use a 27-day persistence model (PS27), which uses the solar wind speed measured on day $d$ to model the speed on day $d+27$. \cite{reiss16} performed a systematic comparison of the three models with the in-situ measured solar wind speed at 1~AU for the time range 2011 to 2014 (excluding ICME periods) and found that all three model performances give reasonable results enabling to reproduce the quiet solar wind conditions.

As described above, the models used for this study are capable of reproducing large-scale features of the continuously emanated solar wind, but do not take into account transient events such as CMEs. By comparing the in-situ measured solar wind speed data at L1 with the model results, we can therefore study the deviations caused by Earth-impacting CMEs. As a measure for the differences between actual in-situ observations and background solar wind, we compute the a) mean error (ME; arithmetic mean between average observational and model results), b) mean absolute error (MAE; arithmetic mean of absolute differences between observations and model results), and c) root mean square error (RMSE; mean squared difference between observations and model results). The ME indicates disproportionate shifts towards positive or negative differences and the RMSE is computed by squaring the differences, hence, gives more weight on large errors. The MAE can be seen as most realistic estimate of the uncertainties between the modeled background solar wind speed and the actually measured solar wind speed.

We would like to note that all algorithms used to conduct the present analysis together with the scripts of the figures are openly accessible online at \url{https://bitbucket.org/reissmar/icme-preconditioning}.

\section{Results}

\begin{figure*}
	\centering
	\includegraphics[width=\textwidth]{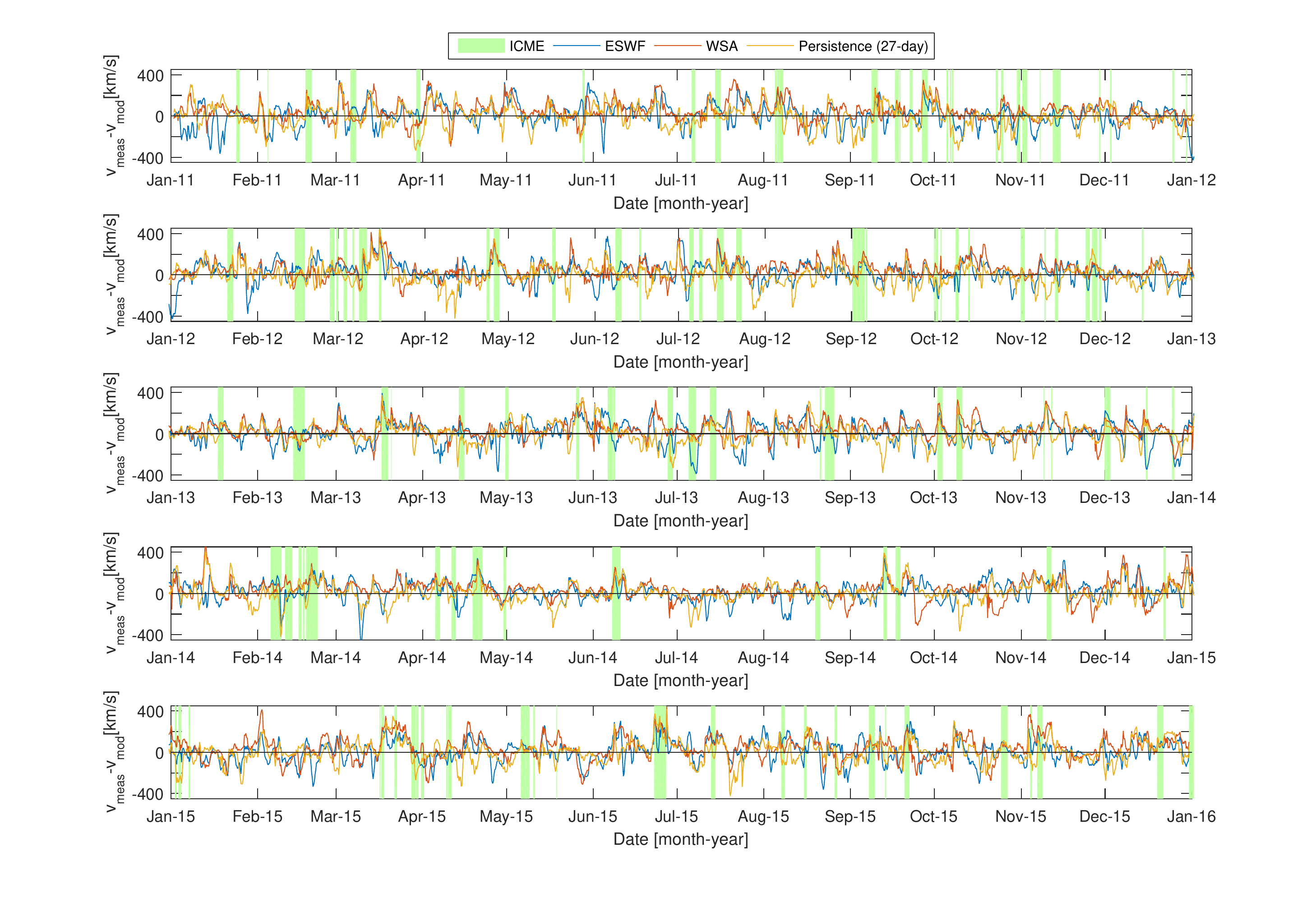}
 	\caption{Differences between the measured solar wind speed (ACE/SWEPAM) and the modeled solar wind speed applying ESWF (blue line), WSA (red line) and PS27 (orange line) during the years 2011--2015. The duration of each recorded ICME ($t_{\rm ICME}$), as given in the Richardson and Cane catalog, is marked as vertical green bar.}
	\label{fig:1}
\end{figure*}

Figure~\ref{fig:1} shows for the entire time range under study (2011--2015) the derived differences between measured solar wind speed (ACE/SWEPAM) and modeled solar wind speed. In total, there are 136 ICMEs listed during 2011--2015, whose duration (ICME shock arrival until end of magnetic structure) is marked with green vertical bars. 22 of the 136 ICMEs arrive simultaneously with high-speed solar wind streams. The mean impact speed of the ICMEs is derived with $430 \pm 76$~km/s (minimum and maximum speed is 290~km/s and 680~km/s, respectively). Inspecting the profiles, we find that CMEs often lead to prolonged speed differences after their transit. In general, PS27 reveals smaller deviations for times without ICMEs compared to ESWF and WSA. 

\begin{figure*}
	\centering
	\includegraphics[width=\textwidth]{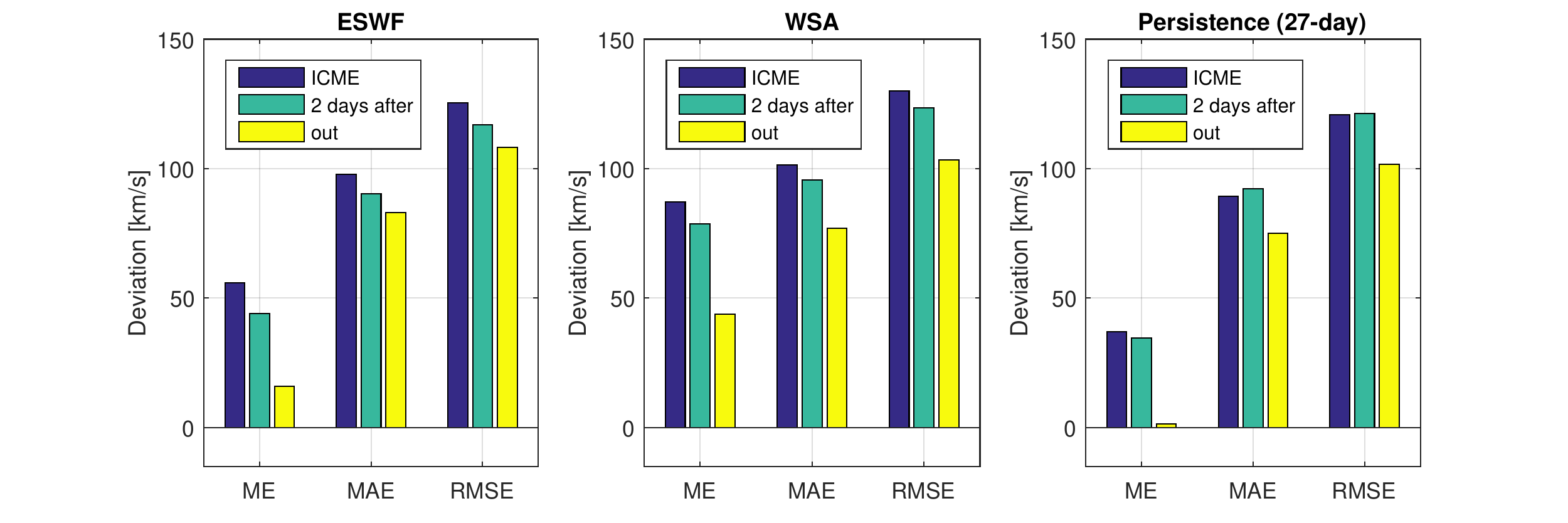}
 	\caption{Differences between measured and modeled (left to right: ESWF, WSA, PS27) solar wind speed calculated by applying the mean error (ME), mean absolute error (MAE), and root mean square error (RMSE). The values cover different selected time intervals, $t_{\rm ICME}$ during the ICME (blue), $t_{\rm +2d}$ 2 days after the end of the ICME magnetic structure (cyan), and $t_{\rm out}$ outside recorded ICME intervals (yellow). The numerical output is given in Table~\ref{error}).}
	\label{fig:2}
\end{figure*}

Figure~\ref{fig:2} gives the errors (ME, MAE, RMSE) describing the deviation between modeled background solar wind speed and observed solar wind speed profiles calculated for three time intervals. Time interval (1) $t_{\rm ICME}$ covers the entire ICME time range, starting with the arrival of the shock-sheath structure until the end of the magnetic cloud, (2) $t_{\rm +2d}$ covers all data points within 2 days starting after the end of the magnetic cloud. (3) $t_{\rm out}$ is for all other time ranges outside ICME occurrences. We find for all error measures larger values during $t_{\rm ICME}$ compared to $t_{\rm out}$. As presumed, this clearly reveals that all background solar wind models are associated with larger errors during the occurrence of transient events. The MAE calculated for the different models over the interval $t_{\rm out}$ lies in the range of 75.1--83.1~km/s, to which we refer to as ``baseline level'' indicating the undisturbed conditions. Compared to this baseline level, the MAE during $t_{\rm ICME}$ is enhanced by 18--32\% and during $t_{\rm +2d}$ by 9--24\%. A similar outcome is derived for the RMSE (cf.\,Table~\ref{error} for all derived results using different models and error measures). From this we conclude that even within 2 days after the magnetic structure of the ICME has passed the spacecraft, the solar wind speed did not go back to the baseline level. This is analyzed in more detail in the following.

\begin{table}[ht]
\centering
\caption{Deviation between observed and modeled solar wind speed: we list the different error measures, mean error (ME), mean absolute error (MAE), and root mean square error (RMSE) given in [km/s], calculated for different models ESWF, WSA, and PS27, covering three time intervals under study, $t_{\rm ICME}$, $t_{\rm +2d}$, and $t_{\rm out}$  (cf.\,Figure~\ref{fig:2}). }
\label{error}
\begin{tabular}{ccrrr}
Model & Error & $t_{\rm ICME}$ & $t_{\rm +2d}$ & $t_{\rm out}$  \\
\hline
ESWF & ME  & 55.9 & 44.1 & 16.0 \\
ESWF & MAE & 97.8 & 90.4 & 83.1 \\
ESWF & RMSE & 125.5 & 117.0 & 108.2 \\
WSA & ME & 87.2 & 78.6 & 43.7 \\
WSA & MAE & 101.5 & 95.5 & 77.1 \\
WSA & RMSE & 130.1 & 123.4 & 103.4 \\
PS27 & ME & 37.0 & 34.7 & 1.3 \\
PS27 & MAE & 89.4 & 92.3 & 75.1 \\
PS27 & RMSE & 120.8 & 121.2 & 101.7 \\
\hline
\end{tabular}
\end{table}

\begin{figure*}
	\centering
	\includegraphics[width=\textwidth]{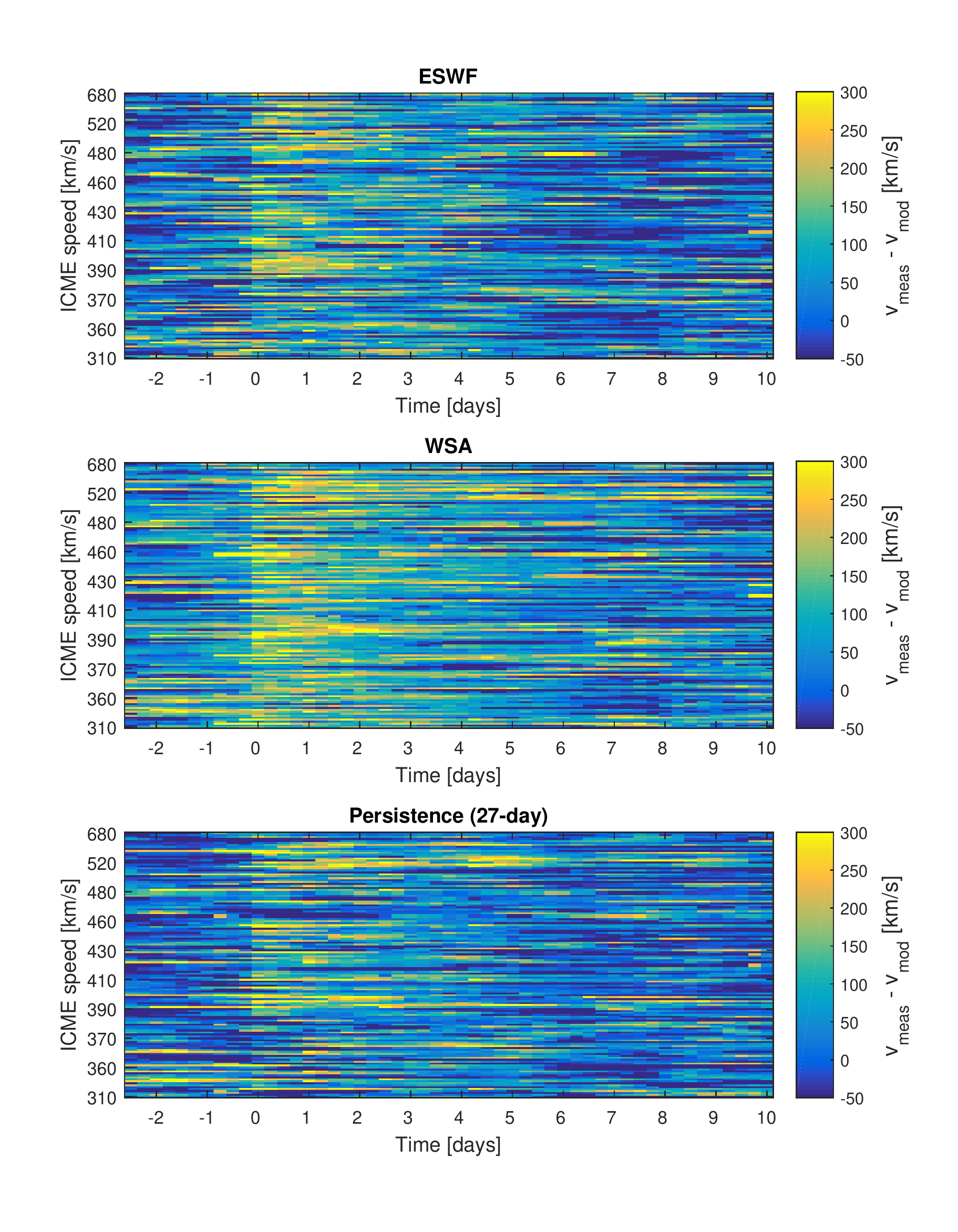}
 	\caption{Speed differences (color coded) for each of the recorded ICMEs stacked by the ICME impact speed as function of time. The differences are calculated from different background solar wind models (top to bottom: ESWF, WSA, PS27). The reference time zero refers to the shock arrival of the ICME.}
	\label{fig:3}
\end{figure*}

We first illustrate the differences between the background solar wind speed and the ICME mean impact speed (see column (i) in the ``R\&C List''), for which we extract for each ICME the period covering 2.5~days before and 10~days after the arrival of the ICME ($t=0$). Figure~\ref{fig:3} shows for each of the 136 ICME events the speed differences between observations and model as function of time (with $t$=0 referring to the arrival time of the ICME shock-sheath structure), stacked in order by the ICME mean impact speed (y-axis top to bottom -- fastest to slowest events). With the start of the ICME impact at the spacecraft, the deviation between modeled and measured speed increases lasting for at least 3~days. From this, no clear relation is obtained between ICME mean impact speed and duration of disturbed conditions. 

\begin{figure*}
	\centering
	\includegraphics[width=\textwidth]{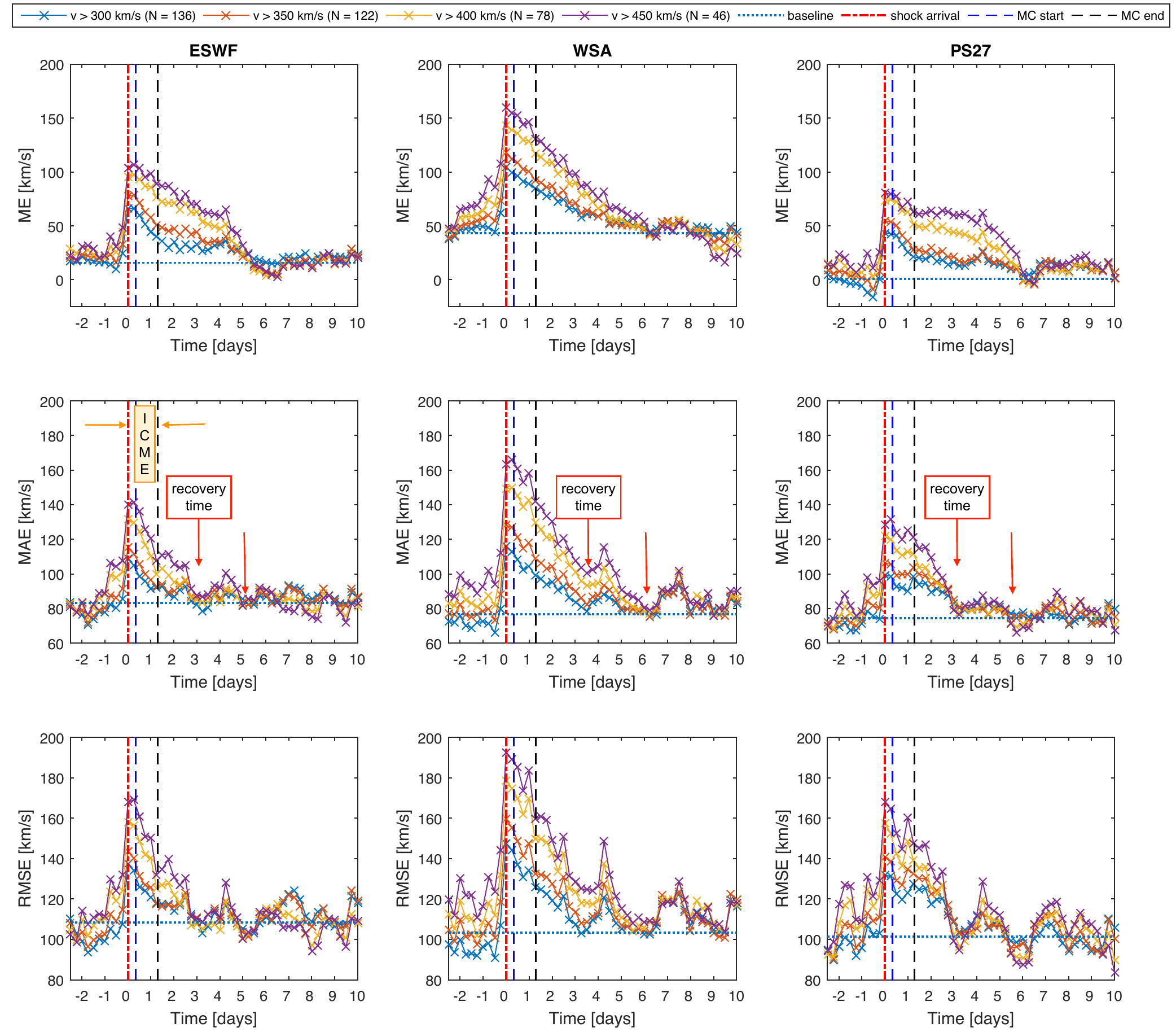}
 	\caption{Computed error measures (ME, MAE, RMSE) for different background solar wind models (ESWF, WSA, PS27). The red dashed-dotted lines indicate the ICME shock arrival, and the blue and black dashed lines indicate the mean start and end time of the magnetic cloud (MC) structure \citep[according to ``R\&C'' list][]{richardson10}. Different colors for the profiles of the calculated error measures indicate the four ICME mean impact speed categories (see legend for speed and sample size). The blue dotted horizontal line is the deviation calculated outside recorded ICME disturbances ($t_{\rm out}$). Focusing on the results from the MAE (middle panels), we mark the mean ICME duration, and the recovery time (red arrows), hence, the time until the increased deviations drop back to the baseline level.}
	\label{fig:4}
\end{figure*}

While Figure~\ref{fig:3} provides a visual assessment of the deviations between modeled and measured speed, Figure~\ref{fig:4} provides the quantification of the required time for the interplanetary medium to recover from transient events. To this end, Figure~\ref{fig:4} shows the computed error measures ME, MAE, RMSE (deviations between observed and modeled solar wind speed) as calculated from the different models (ESWF, WSA, PS27) covering the period from 2.5~days before to 10~days after the arrival of the ICME ($t=0$). We give the results for four different ICME categories which accord to different ICME mean impact speeds. To ensure a sufficiently large sample size, the four categories cover 136 ICMEs with $v>300$km/s, 122 ICMEs with $v>350$ km/s, 78 ICMEs with $v>400$, and 46 ICMEs with $v>450$. The reference time zero is the shock arrival time of each ICME at the in-situ spacecraft. Hence, the derived similarities in the error profiles of the four speed categories reflect a characteristic behavior of the evolution of the disturbances. As indication of the undisturbed conditions, the calculated deviations outside recorded ICME disturbances ($t_{\rm out}$), hence, baseline level for the MAE results, are given by blue dotted horizontal lines. The ICME shock-sheath and magnetic structure is marked by vertical red dashed-dotted and blue/black dashed lines. The listed events under study have a mean duration for the entire ICME, $t_{\rm ICME}$, of 1.27$\pm$0.67~days and a mean duration for the magnetic structure of 0.96$\pm$0.53~days (calculated from the ``R\&C List''). Inspecting the error profiles for the entire sample ($v>300$~km/s), all models reveal the largest deviations during the passage time of the ICME shock-sheath structure, followed by a rather linear decrease of the error profile during the passage of the magnetic cloud structure. From ESWF and WSA models and for each ICME category, we derive that all computed error measures continue to be higher than the baseline level and only start to decrease at about $t_0+3$~days lasting until about $t_0+6$~days (for faster events with $v>400$~km/s). This recovery time is much longer than the average duration of the entire disturbance. Independent from the ICME category, measure errors and models used, a short-term enhancement in the deviations around 4--5~days is revealed. The higher the mean impact speed of the ICME event the larger the derived values for the ME, MAE and RMSE. We note that with the smaller subset for faster ICMEs, larger fluctuations in the errors before the shock arrival are derived, which is related to the smaller statistics.

\section{Discussion and conclusions}

In this study we evaluate the impact of ICMEs on the solar wind conditions in interplanetary space for the time range 2011--2015. To simulate the quiet solar wind speed we use different (semi-)empirical models such as the ESWF, WSA, and PS27. For periods affected by ICMEs we find clear deviations that last at least 3~days (up to 6~days) after the ICME start. This is longer than the duration of the ICME disturbance itself (for our sample on average $\sim$1.3~days) as it passes the in-situ spacecraft. Prolonged geomagnetic storms and long-duration magnetic cloud structures as a consequence of ICME flank hits are found to last $\sim$1.5--2~days \citep{marubashi07,moestl10}. Although the investigated approach introduces uncertainties due to the analysis of stacked events, our results for the first time quantify over which time interval ICME disturbances affect/precondition interplanetary space (total disturbance period is $\sim$3--6~days) and how long interplanetary space may need to recover from that (total disturbance period minus ICME duration is $\sim$2--5~days). With these results, we can confirm conjectured scenarios why the CME from July 23, 2012 became super-fast \citep{liu14,temmer15}. Besides the solar wind speed, we presume that also other solar wind properties as the magnetic field and the density are affected over a similar time span. 

There is some indication that faster ICMEs might disrupt the interplanetary medium for a longer time span compared to slow ICMEs. However, the derived fluctuations are found to be larger for higher ICME impact speeds. We stress that the time range under study covers the weak solar cycle 24, hence, very fast ICME events are actually missing in this sample. We also note that a fraction of about 15\% of the listed ICMEs are related to enhanced solar wind speed streams emanating from coronal holes \citep[relation of stealth ICMEs to coronal holes see e.g.,][]{dhuys14}. For those events we derived no significant deviations between modeled and observed solar wind speed.  

The question remains what causes such specific long-lasting preconditioning periods? (1) The time interval is comparable to the average solar wind transit time from Sun to Earth, and might reflect the time span that the entire system, i.e.\,the interplanetary medium, needs to recover from the propagating disturbance. (2) The occurrence of multiple ICMEs, overlapping with each other, might cause prolonged disturbances. We cross-checked this possibility for a subset of 57 events, for which all ICMEs that had other ICMEs surrounding them with $\sim$2.5~days were removed, and performed the analysis again. We still found the same trend of prolonged preconditioning. This may indicate that (3) other kind of disturbances, or non-listed ICMES, traveling in the wake of the listed ICMEs might cause this effect (the derived short-term increase in the deviations around 4--5~days might support this scenario). Aftermath regions of ICMEs without clear rotating magnetic field structures could be due to reconnection of the magnetic structure with the interplanetary magnetic field \citep{dasso07}. Peaks in plasma density and speed observed close to the trailing edge of magnetic flux ropes were interpreted as compression caused by high-speed solar wind streams running into the expanding ICME \citep{rodriguez16}.

The obtained results have strong implications for studying CME propagation behavior and also for space weather forecasting, as they provide quantitative information on the reliability of operational forecast models during ICME disturbances.

\acknowledgments
We thank the anonymous reviewer for helpful comments. M.~T. and M.~R. gratefully acknowledge the NAWI Graz funding \textit{F\"orderung von JungforscherInnengruppen 2013--2015}. We greatly acknowledge the Austrian Science Fund FWF: V195-N16 and P24092-N16. L.~N. performed this work as part of Natural Resources Canada's Public Safety Geoscience program.

\bibliographystyle{aasjournal}

\newpage

\end{document}